\documentclass[10pt, twocolumn]{IEEEtran}

\usepackage{amsmath, mathrsfs,amsfonts}
\usepackage{amsfonts}
\usepackage{amssymb}
\usepackage{acronym}
\usepackage{graphicx}

\def\ignore#1{}

\def\bs{\boldsymbol}

\def\foral{\textrm{for all }}

\newtheorem{dom_face}{Definition}
\newtheorem{expansion}[dom_face]{Definition}
\newtheorem{Hausdorff}[dom_face]{Definition}

\newtheorem{rate-splitting}{Theorem}
\newtheorem{mod-algorithm}[rate-splitting]{Theorem}
\newtheorem{convergence-thm}[rate-splitting]{Theorem}

\newtheorem{Bound2}[rate-splitting]{Theorem}
\newtheorem{Bound1}[rate-splitting]{Theorem}

\newtheorem{half-space-proj}{Lemma}
\newtheorem{psuedo-nonexp}[half-space-proj]{Lemma}

\newtheorem{region_chebyshev}[half-space-proj]{Lemma}
\newtheorem{opt_dist}[half-space-proj]{Lemma}

               {\begin{list}{}{\leftmargin#1\rightmargin#2\topsep#3}\item{}}%
               {\end{list}}

\begin{document}
\title{Rate and Power Allocation in Fading Multiple Access Channels}


\author{Ali ParandehGheibi\thanks{A.\ ParandehGheibi is with the Laboratory for
Information and Decision Systems, Electrical Engineering and Computer Science Department,
Massachusetts Institute of Technology, Cambridge MA, 02139 (e-mail: parandeh@mit.edu)}, Atilla
Eryilmaz\thanks{A.\ Eryilmaz is with the Electrical and Computer Engineering, Ohio State
University, OH, 43210 (e-mail: eryilmaz@ece.osu.edu)}, Asuman Ozdaglar, and Muriel M\'edard\thanks{
A.\ Ozdaglar and M.\ M\'edard are with the Laboratory for Information and Decision Systems,
Electrical Engineering and Computer Science Department, Massachusetts Institute of Technology,
Cambridge MA, 02139 (e-mails: asuman@mit.edu, medard@mit.edu)}}


\maketitle
\thispagestyle{headings}

\begin{abstract}
We consider the problem of rate and power allocation in a fading multiple-access channel. Our
objective is to obtain rate and power allocation policies that maximize a utility function defined
over average transmission rates. In contrast with the literature, which focuses on the linear case,
we present results for general concave utility functions. We consider two cases. In the first case,
we assume that power control is possible and channel statistics are known. In this case, we show
that the optimal policies can be obtained greedily by maximizing a linear utility function at each
channel state. In the second case, we assume that power control is not possible and channel
statistics are not available. In this case, we define a greedy rate allocation policy and provide
upper bounds on the performance difference between the optimal and the greedy policy. Our bounds
highlight the dependence of the performance difference on the channel variations and the structure
of the utility function.
\end{abstract}

\section{Introduction}
Dynamic allocation of communication resources such as bandwidth or transmission power is a central
issue in multiple access channels in view of the time varying nature of the channel and
interference effect. Most of the existing literature focuses on specific communication schemes such
as TDMA (time-division multiple access) \cite{TDMA} and CDMA (code-division multiple access)
\cite{CDMA1,CDMA3} systems. An exception is the work by Tse \emph{et al.} \cite{Tse}, who consider
the notion of \emph{throughput capacity} for the fading channel with Channel State Information
(CSI). That is the notion of Shannon capacity applied to the fading channel, where the codeword
length can be arbitrarily long to average over the fading of the channel. The points on the
boundary of the capacity region are attained by dynamically allocating the resources with the goal
of maximizing a \emph{linear} utility function.

In this paper, we consider the problem of rate and power allocation in a multiple access channel
with perfect CSI. Contrary to the linear case in \cite{Tse}, we consider maximizing a general
utility function of transmission rates over the throughput capacity region. Such a general concave
utility function allows us to capture different performance metrics such as fairness or delay (c.f.
Shenker \cite{She95}, Srikant \cite{Srikant}). Our contributions can be summarized as follows.

We first consider the case where channel statistics are known and power can be controlled at the
transmitters. Owing to strict convexity of the capacity region, we show that the resource
allocation problem for a general concave utility  is equivalent to another problem with a linear
utility. Hence, the optimal resource allocation policies are obtained by applying the results in
\cite{Tse} for the linear utility. Given a general utility function, the conditional gradient
method is used to obtain the corresponding linear utility. Second, we consider the case where the
transmitters do not have the power control feature and channel statistics are not known. In this
case, a greedy policy which maximizes the utility function for any given channel state, is
suboptimal. However, we can bound the performance difference between the optimal and the greedy
policies. We show that this bound is tight in the sense that it goes to zero either as the utility
function tends to a linear function of the rates or as the channel variations vanish.

Other than the papers cited above, our work is also related to the work of Vishwanath \emph{et al.}
\cite{Vishwanath} which builds on \cite{Tse} and takes a similar approach to the resource
allocation problem for linear utility functions. Other works address different criteria for
resource allocation including minimizing the weighted sum of transmission powers \cite{power_min},
and considering Quality of Service (QoS) constraints \cite{QoS}. In contrast to this literature, we
consider the utility maximization framework for general concave utility functions.

The remainder of this paper is organized as follows: In Section II, we introduce the model and
describe the capacity region of a fading multiple-access channel. In Section III, we address the
resource allocation problem with power control and known channel statistics. In Section IV, we
consider the same problem without power control and channel statistics. Finally, we give our
concluding remarks in Section V.

Regarding the notation, we denote by $x_i$ the $i$-th component of a vector $\bs x$. A vector $\bs
x$ is positive when $x_i> 0$ for all components $i$ of $\bs x$. We denote the nonnegative orthant
by $\mathbb{R}^n_+$, i.e., $\mathbb{R}^n_+ = \{\bs x\in \mathbb{R}^n\mid \bs x\ge 0\}$. We write
$\bs x'$ to denote the transpose of a vector $\bs x$.

\section{System Model}
We consider $M$ transmitters sharing the same media to communicate to a single receiver. We model
the channel as a Gaussian multiple access channel with flat fading effects
\begin{equation}\label{fading_model}
    Y(n) = \sum_{i=1}^M \sqrt{H_i(n)} X_i(n) + Z(n),
\end{equation}
where $X_i(n)$ and $H_i(n)$ are the transmitted waveform and the fading process of the
\textit{i}-th transmitter, respectively, and $Z(n)$ is white Gaussian noise with variance $N_0$. We
assume that the fading processes of all transmitters are jointly stationary and ergodic, and the
stationary distribution of the fading process has continuous density. We also assume that all the
transmitters and the receiver have instant access to channel state information. In practice, the
receiver measures the channels and feeds back the channel information to the transmitters. The
implicit assumption in this model is that the channel variations are much slower than the data
rate, so that the channel can be measured accurately at the receiver and the amount of feedback
bits is negligible compared to that of transmitting information.

First, consider the non-fading case where the channel gains are fixed. The capacity region of the
Gaussian multiple-access channel with no power control is described as follows \cite{cover}
\begin{eqnarray}\label{Cg}
    C_g(\bs P, \bs h) &=& \bigg\{ \bs R \in \mathbb{R}^M_+:
    \sum_{i \in S} R_i \leq  C\Big(\sum_{i \in S} h_i P_i, N_0\Big), \nonumber \\
    && \quad \textrm{for all}\  S \subseteq \mathcal M = \{1, \ldots, M\} \bigg\},
\end{eqnarray}
where $P_i$ and $R_i$ are the \emph{i}-th transmitter's power and rate, respectively. $C(P,N)$
denotes Shannon's formula for the capacity of AWGN channel given by
\begin{equation}\label{C_AWGN}
    C(P,N) = \frac{1}{2}\log(1+\frac{P}{N}) \quad \textrm{nats}.
\end{equation}

For a multiple-access channel with fading, but fixed transmission powers $P_i$, the
\emph{throughput} capacity region is given by averaging the instantaneous capacity regions with
respect to the fading process \cite{Shamai},
\begin{eqnarray}\label{Ca}
    C_a(\bs P) &=& \bigg\{ \bs R \in \mathbb{R}^M_+: \sum_{i \in S} R_i
    \leq \mathbb{E}_{\bs H} \bigg[ C\Big(\sum_{i \in S} H_i P_i, N_0\Big) \bigg], \nonumber \\
    && \qquad \qquad \qquad \qquad \textrm{for all} \  S  \subseteq \mathcal M \bigg\},
\end{eqnarray}
where $\bs H$ is a random vector with the stationary distribution of the fading process.

A power control policy $\mathcal{P}$ is a map from any given fading state $\bs h$ to
$\mathcal{P}(\bs h) =(\mathcal{P}_1(\bs h), \ldots, \mathcal{P}_M(\bs h))$, the powers allocated to
the transmitters. Similarly, we can define the rate allocation policy, $\mathcal R$, as a map from
the fading state $\bs h$ to the transmission rates, $\mathcal R(\bs h)$. For any given
power-control policy $\mathcal{P}$, the capacity region follows from (\ref{Ca}) as
\begin{eqnarray}\label{Cf}
    C_f(\mathcal{P}) &=& \bigg\{\bs R \in \mathbb{R}^M_+: \sum_{i \in S} R_i \leq \nonumber \\ &&
    \mathbb{E}_{\bs H} \bigg[ C\Big(\sum_{i \in S} H_i \mathcal{P}_i(\bs H),
    N_0\Big) \bigg], \textrm {for all}\  S \subseteq \mathcal M \bigg\}. \nonumber
\end{eqnarray}
Tse \emph{et al.} \cite{Tse} have shown that the throughput capacity of a multiple access fading channel is given by 
\begin{equation}\label{C_power_ctrl}
    C(\bar{\bs P}) = \bigcup_{\mathcal{P} \in \mathcal{G}} C_f(\mathcal{P}),
\end{equation}
where $\mathcal{G} = \{ \mathcal{P}: \mathbb{E}_{\bs H} [\mathcal{P}_i(\bs H)] \leq \bar{P}_i,
\textrm{for all}\ i\} $ is the set of all power control policies satisfying the average power
constraint. Let us define the notion of boundary or dominant face for any of the capacity regions
defined above.
\begin{dom_face}\label{dom_face}
The \emph{dominant face} or \emph{boundary} of a capacity region, denoted by $\mathcal{F}(\cdot)$,
is defined as the set of all $M$-tuples in the capacity region such that no component can be
increased without decreasing others while remaining in the capacity region.
\end{dom_face}

\section{Rate Allocation with Power Control}

        In this section, we assume that the channel statistics are known a priori. The goal of
        optimal resource allocation is to find feasible rate and power allocation policies denoted
        by $\mathcal{R}^*$ and $\mathcal{P}^*$, respectively, such that $\mathcal{R}^*(\bs H) \in C_g\big(\mathcal{P}^*(\bs H),\bs
        H\big)$, and $\mathcal{P}^* \in \mathcal G$. Moreover,
    \begin{eqnarray}\label{RAC_pctrl}
       \mathbb{E}_{\boldsymbol{H}} [\mathcal{R}^*(\bs H)] = &\bs R^* =& \textrm{argmax} \quad u(\bs R) \nonumber \\ && \ \textrm{subject to} \quad \bs R \in  C(\bar{\bs P})
      \end{eqnarray}
        where $u(\cdot)$ is a given utility function and is assumed to be a continuously differentiable concave function of $\bs R$, and
        monotonically increasing in each component $R_i$ for all $i$.

        For the case of a linear utility function, i.e., $u(\boldsymbol R) = \boldsymbol \mu '
    \boldsymbol R $ for some $\boldsymbol \mu \in \mathbb{R} _{+}^M $, Tse \emph{et al.} \cite{Tse}
    have  shown that the optimal rate and power allocation policies are given by the optimal solution to a
    linear program, i.e.,
        \begin{eqnarray}\label{LP_RAC_pctrl}
            \left(\mathcal R^*(\bs h), \mathcal P^*(\bs h)\right) &=& \textrm{arg}\max_{\boldsymbol r , \boldsymbol p} \left( \boldsymbol \mu ' \boldsymbol r - \boldsymbol
            \lambda ' \boldsymbol p \right) \nonumber \\ &&\ \textrm{subject to} \quad \boldsymbol r \in
            C_g(\boldsymbol h, \boldsymbol p),
        \end{eqnarray}
        where $\boldsymbol h$ is the channel state realization, and $\boldsymbol \lambda \in \mathbb{R}
        _{+}^M$ is a Lagrange multiplier satisfying the average power constraint, i.e.,
        $\bs \lambda$ is the unique solution of the following equations

        \begin{eqnarray}\label{lambda_mu}
            &&\!\!\!\!\!\!\!\!\!\!\!\!\!\!\!\!\int_0^{\infty}\!\! \frac{1}{h} \int_{\frac{2
            \lambda_i (N_0+z) }{\mu_i}}^{\infty} \nonumber \\ &&\!\!\!\!\!\! \prod_{k \neq i} F_k \left( \frac{2 \lambda_k h
            (N_0 + z)}{2 \lambda_i (N_0+z) + (\mu_k - \mu_i) h} \right) f_i(h) \mathrm{d}h
            \mathrm{d}z = \bar{P}_i \nonumber \\
        \end{eqnarray}

        where $F_k$ and $f_k$ are cumulative distribution function (CDF) and probability density
        function (PDF) of the stationary distribution of the channel state process for transmitter $k$,
        respectively.

        Exploiting the polymatroid structure of the capacity region, problem (\ref{LP_RAC_pctrl}) can be
        solved by a simple greedy algorithm (see Lemma 3.2 of \cite{Tse}). It is also shown in \cite{Tse}
        that for positive $\boldsymbol \mu$ the optimal solution, $\boldsymbol R^*$, to the problem in
        (\ref{RAC_pctrl}) is \emph{uniquely} obtained. Given the distribution of channel state process,
        denoted by $F_k$ and $f_k$, we have
        \begin{eqnarray}\label{R_mu}
            R_i^*(\boldsymbol \mu) &=& \int_0^{\infty}\!\!\!\! \frac{1}{2(N_0+z)} \int_{\frac{2
            \lambda_i (N_0+z) }{\mu_i}}^{\infty} \nonumber \\
            && \!\!\!\!\!\!\!\prod_{k \neq i} F_k \left( \frac{2 \lambda_k h
            (N_0 + z)}{2 \lambda_i (N_0+z) + (\mu_k - \mu_i) h} \right) f_i(h) \mathrm{d}h
            \mathrm{d}z, \nonumber \\
        \end{eqnarray}
         The uniqueness of $\bs R^*$ follows from the fact that the stationary
        distribution of the fading process has continuous density \cite{Tse}. It is worth
        mentioning that (\ref{R_mu}) parametrically describes the \emph{boundary} of the capacity
        region, and hence, there is a one-to-one correspondence between the boundary of
        $C(\bs{\bar{P}})$ and the positive vectors $\bs \mu$ with unit norm.

        Now consider a general concave utility function. We use an iterative method to compute the optimal solution, $\bs R^*$, of problem
        (\ref{RAC_pctrl}). Note that by monotonicity of the utility function, $\bs R^*$ always lies
        on the \emph{boundary} of the capacity region, $C(\bs{\bar{P}})$. Once $\bs R^*$ is known,
        then in view of one-to-one correspondence between the boundary of
        $C(\bs{\bar{P}})$ and the positive vectors $\bs \mu$, there exist a positive vector $\bs
        \mu^*$ such that
        \begin{equation}\label{R_mu_corresp}
            \bs R^* = \textrm{argmax} \quad (\bs\mu ^*)' \bs R \quad \textrm{subject to} \quad \bs R \in
            C(\bar{\bs P}).
        \end{equation}

        Therefore the optimal rate and power allocation policies can be obtained by using the greedy
        policies of Tse \emph{et al.} \cite{Tse} for the linear utility function, $u(\bs R) = (\bs\mu ^*)' \bs
        R$.

        We use the conditional gradient method \cite{nlp} in order to iteratively compute the optimal
        solution of problem (\ref{RAC_pctrl}). The $k$-th iteration of the method is given by
        \begin{equation}\label{frank-wolfe}
            \bs R^{k+1} = \bs R^k + \alpha^k(\bs{\bar{R}}^k - \bs R^k),
        \end{equation}
        where $\alpha^k $ is the stepsize and $\bs{\bar{R}}^k$ is obtained as
        \begin{equation}\label{frank-wolfe2}
            \bs{\bar{R}}^k \in \textrm{arg}\!\!\!\!\max_{\bs{R} \in C(\bs{\bar{P}})} \left( \nabla u(\bs{R}^k)'(\bs
            R - \bs R^k)\right).
        \end{equation}
        Since the utility function is monotonically increasing, the gradient vector is always
        positive and, hence, the unique optimal solution to the above sub-problem is obtained by
        (\ref{R_mu}), in which $\bs \mu$ is replaced by $\nabla u(\bs{R}^k)$. By concavity of the
        utility function and convexity of the capacity region, the iteration (\ref{frank-wolfe})
        will converge to the optimal solution of (\ref{RAC_pctrl}) for appropriate stepsize
        selection rules such as Armijo rule or limited maximization rule (c.f. \cite{nlp} pp. 220-222).

        Note that our goal is to determine rate and power allocation policies. Finding $\bs R^*$
        allows us to determine such policies by the greedy policy in  (\ref{LP_RAC_pctrl}) for $\bs \mu
        ^* = \nabla u(\bs{R}^*)$. It is worth mentioning that all the computations for obtaining $\bs
        R^*$ are performed once in the setup of the communication session. So the convergence rate of the conditional
        gradient method is generally not of critical importance.

\section{Rate Allocation without Power Control}
        In this section we assume that the channel statistics are not known and that the
        transmission powers are fixed to $\bs P$. In practice, this scenario occurs when the
        transmission power may be limited owing to environmental limitations such as human presence,
        or limitations of the hardware.

        The capacity region of the multiple access channel for this scenario is a polyhedron and is
        given by (\ref{Ca}). Similarly to the previous case, the optimal rate allocation
        policy, $\mathcal{R}^*(\cdot)$, is such that $\mathcal{R}^*(\bs H) \in C_g(\bs P, \bs H)$, and
        \begin{eqnarray}\label{RAC_npctrl}
            \mathbb{E}_{\boldsymbol{H}} [\mathcal{R}^*(\bs H)] = &\bs R^* &\in \textrm{argmax} \quad u(\bs R)\nonumber \\ && \quad \textrm{subject to} \quad \bs R \in  C_a(\bs
            P).
        \end{eqnarray}
        It is worth mentioning that the approach used to find the optimal resource allocation
        policies for the previous case does not apply to this scenario, because $C_g(\bs
        P, \bs h)$ is a polyhedron and hence, the uniqueness property of $\bs R^*$ for any positive vector
        $\bs \mu$ does not hold anymore.

        Here we present a \emph{greedy} rate allocation policy and compare its performance with the
        unknown optimal policy. The performance of a particular rate allocation policy is defined
        as the utility at the average rate achieved by that policy. The greedy policy, denoted by
        $\bar{\mathcal{R}}(\cdot)$, optimizes the utility function for any channel realization.
        i.e.,
        \begin{equation}\label{R_greedy}
            \bar{\mathcal{R}}(\bs h) = \textrm{argmax}_{\bs R \in C_g(\bs P, \bs h)} \quad
            u(\bs R).
        \end{equation}
        Consider the following relations
        \begin{eqnarray}\label{jensen}
          \mathbb{E}_{\bs H}\big[u\big(\mathcal{R}^*(\bs H)\big)\big] &\leq& \mathbb{E}_{\bs H}\big[u\big(\bar{\mathcal{R}}(\bs H)\big)\big] \nonumber \\
          &\leq& u\big(\mathbb{E}_{\bs H}\big[\bar{\mathcal{R}}(\bs H)\big]\big) \nonumber \\
          &\leq& u\big(\mathbb{E}_{\bs H}\big[\mathcal{R}^*(\bs H)\big]\big),
        \end{eqnarray}
        where the second inequality follows from the Jensen's inequality by concavity of the utility function.

        In the case of a linear utility function we have $u\big(\mathbb{E}_{\bs H}\big[\mathcal{R}^*(\bs
        H)\big]\big) =  \mathbb{E}_{\bs H}\big[u\big(\mathcal{R}^*(\bs H)\big)\big]$, so equality holds throughout in (\ref{jensen}) and
        $\bar{\mathcal{R}}(\cdot)$ is indeed the optimal rate allocation policy. For
        nonlinear utility functions, the greedy policy can be strictly suboptimal.

        However, the greedy policy is not arbitrarily worse than the optimal one. In view of (\ref{jensen}), we can bound the
        performance difference, $u(\bs{R}^*) - u\big(\mathbb{E}_{\bs H}\big[\bar{\mathcal{R}}(\bs H)\big]\big)$, by
        bounding $\Big|u\big(\mathbb{E}_{\bs H}\big[\mathcal{R}^*(\bs H)\big]\big) - u\big(\mathbb{E}_{\bs H}\big[\bar{\mathcal{R}}(\bs H)\big]\big)\Big|$ or
        $\Big|u\big(\mathbb{E}_{\bs H}\big[\mathcal{R}^*(\bs H)\big]\big) -  \mathbb{E}_{\bs H}\big[u\big(\mathcal{R}^*(\bs
        H)\big)\big]\Big|$ from above. We show that the first bound goes to zero as the channel variations
        become small and the second bound vanishes as the utility function tends to have a more
        linear structure.

        Before stating the main theorems, let us introduce some useful definitions and lemmas.

        \begin{expansion}\label{expansion_def}
        Let $Q$ be a polyhedron described by a set of linear constraints, i.e.,
        \begin{equation}\label{polyhedron}
            Q = \left\{\bs x \in \mathbb{R}^n: A \bs x \leq \bs b \right\}.
        \end{equation}
        Define the \emph{expansion} of $Q$ by $\delta$, denoted by $\mathcal{E}_\delta(Q)$, as the polyhedron
        obtained by relaxing all the constraints in (\ref{polyhedron}), i.e., $ \mathcal{E}_\delta(Q) = \left\{\bs x \in \mathbb{R}^n: A \bs x \leq \bs b + \delta\mathbf{1}
            \right\},$
        where $\mathbf{1}$ is the vector of all ones.


        \end{expansion}

        \begin{Hausdorff}\label{Hausdorff_def}
        Let $X$ and $Y$ be two polyhedra described by a set of linear constraints. Let
        $\mathcal{E}_d(X)$ be an \emph{expansion} of $X$ by relaxing its constraints by $d$. The distance
        $d_H(X,Y)$ between $X$ and $Y$ is defined as the minimum scalar $d$ such that $X \subseteq \mathcal{E}_d(Y)$ and $ Y \subseteq
        \mathcal{E}_d(X)$.
        \end{Hausdorff}

       Lemma \ref{region_chebyshev} extends
        Chebychev's inequality for capacity regions. It states that the time varying capacity
        region does not deviate much from its mean with high probability.
        \begin{region_chebyshev} \label{region_chebyshev}
            Let $\bs H$ be a random vector with the stationary distribution of the fading process
            with mean $\bs{\bar{H}}$ and covariance matrix $K$. Then
            \begin{equation}\label{capacity_cheby}
                \textbf{\textrm{Pr}} \Big\{ d_H \left(C_g(\bs{P},\bs{H}), C_a(\bs{P}) \right) > \delta \Big\}
                \leq \frac{\sigma_H^2}{\delta^2},
            \end{equation}
            where $\sigma_H^2$ is defined as
              \begin{eqnarray}\label{sigma_H}
                          && \sigma_H^2 \triangleq \frac{1}{4}\sum_{S \subseteq \{1,\ldots, M\}} \bs{\Gamma}_S' K \bs \Gamma_S \Bigg(1+  \nonumber \\
                           && \left[(1+\bs{\Gamma}'_S \bs{\bar{H}})(\sqrt{2 \log(1+\bs{\Gamma}'_S \bs{\bar{H}})} -
                             \frac{\sqrt{\bs \Gamma_S' K \bs \Gamma_S}}{2})\right]^2\Bigg), \nonumber \\
              \end{eqnarray}
                where
                  \begin{equation}\label{P_indicator}
                   {(\bs \Gamma_S)}_i = \left\{ \begin{array}{ll}
                   \frac{P_i}{N_0}, & \textrm{$i \in S$}\\
                   0, & \textrm{otherwise.}
                   \end{array} \right.
                 \end{equation}

        \end{region_chebyshev}

\begin{proof} Define random variables $Y_S$ and $Z_S$ as the following:
                \begin{equation}\label{Y_S}
                 Y_S = \frac{1}{2} \log\big(1+\sum_{i \in S}\frac{H_i P_i}{N_0}\big) = \frac{1}{2} \log(1+Z_S), \quad \textrm{for all}\ S \subseteq
                 \mathcal M.
                \end{equation}
                The facet defining constraints of $C_g(\bs P, \bs H)$ and $C_a(\bs P)$ are of the form of $\sum_{i \in S}R_i \leq
                Y_S$ and $\sum_{i \in S}R_i \leq \mathbb{E}[Y_S]$, respectively. Hence, by
                Definition \ref{Hausdorff_def}, we have $d_H \left(C_g(\bs{P},\bs{H}), C_a(\bs{P}) \right) >
                \delta$ if and only if $|Y_S - \mathbb{E}[Y_S]| > \delta$, for all $S \subseteq \mathcal M$ . After some manipulations, the following
                relations can be verified by employing Chebyshev's
                inequality:
                \begin{eqnarray}
                   \textbf{\textrm{Pr}} \Big\{ d_H \left(C_g(\bs{P},\bs{H}), C_a(\bs{P}) \right) > \delta \Big\} && \nonumber
                \end{eqnarray}
                \begin{eqnarray}\label{chebyshev_pf_eqn}
                     &=& \ \textbf{Pr}\Big\{ \max_{S} |Y_S - \mathbb{E}[Y_S]| > \delta \Big\}    \nonumber \\
                    &\leq& \frac{1}{\delta^2} \sum_{S \subseteq \mathcal M} \sigma^2_{Y_S}
                  \end{eqnarray}
                  where $\sigma^2_{Y_S}$ denotes variance of $Y_s$, and can be bounded from above
                  as follows (c.f. Appendix II, \cite{tech_report})
                  \begin{equation}\label{sigma_Y}
    \sigma^2_{Y_S} \leq \frac{\sigma_{Z_S}^2}{4} \left(1+ \left[(1+\bar{Z_S})(\sqrt{2 \log(1+\bar{Z_S})} -
    \frac{\sigma_{Z_S}}{2})\right]^2\right),
    \end{equation}
where
$$ \bar{Z}_S = \mathbb{E}\Big[ \sum_{i \in S}H_i P_i\Big] =  \bs \Gamma'_S \bs{\bar{H}}, $$
$$ \sigma_{Z_S}^2 = \textrm{var}\Big( \sum_{i \in S}H_i P_i\Big) = \bs\Gamma_S' K \bs \Gamma_S. $$

The desired result is concluded by substituting $\bar{Z_S}$ and $\sigma_{Z_S}^{2}$ in
(\ref{sigma_Y}) and combing the result with (\ref{chebyshev_pf_eqn}).
\end{proof}

        The system parameter $\sigma_H^2$ in Lemma \ref{region_chebyshev} is proportional to
        channel variations, and we expect it to vanish for small channel variations. The
        following lemma ensures that the distance between the optimal solutions of the utility
        maximization problem over two regions is small, provided that the regions are close to each
        other.
        \begin{opt_dist}\label{opt_dist}
             Let $\bs R_1^*$ and $\bs R_2^*$ be the optimal solution of maximizing the utility over $C_g(\bs P,
             \bs H_1)$ and $C_g(\bs P, \bs H_2)$, respectively. If there exist positive scalars $A$ and $B$ such
             that
             \begin{eqnarray}\label{AB_hyp}
             |u(\bs R_1) - u(\bs R_2)| &\leq& B \|\bs R_1- \bs R_2\|, \nonumber \\
             &\foral& \bs R_i \in \mathcal{F}(C_g(\bs P, \bs H_i)) , \quad i=1,2. \nonumber \\
             |u(\bs R_i^*) - u(\bs R_i)| &\geq& A \|\bs R_i^* - \bs R_i\|^2, \nonumber \\
             &\foral& \bs R_i \in C_g(\bs P, \bs H_i), \quad i=1,2, \nonumber \\
             \end{eqnarray}
             and moreover if
             $$d_H\big(C_g(\bs P, \bs H_1), C_g(\bs P, \bs H_2)\big) \leq \delta$$
             then, we have
             \begin{equation}\label{opt_dist_result}
                 \|\bs R_1^* - \bs R_2^*\| \leq
                 {\delta}^{\frac{1}{2}}\left[{\delta}^{\frac{1}{2}}+\Big(\frac{B}{A}\Big)^{\frac{1}{2}}\right].
             \end{equation}
       \end{opt_dist}

        \begin{proof}
                Without loss of generality assume that $u(\bs R_2^*) \geq u(\bs R_1^*)$. To
                simplify the notations for capacity regions, let $C_1 = C_g\big(\bs P, \bs H_1\big)$ be a \emph{polymatroid}, i.e.,
            \begin{equation}\label{polymatroid}
                C_1 = \bigg\{ \bs R \in \mathbb{R}^M_+: \sum_{i \in S} R_i \leq f(S),\
                \textrm{for all}\ S \subseteq \mathcal M \bigg\},
            \end{equation}
             for some submodular function $f(S)$, and let $C_2$ be an \emph{expansion} of $C_1$ by
            $\delta$ as defined in Definition \ref{expansion_def}. We first show that for every $\bs R \in \mathcal{F}(C_2)$, there exists a vector
                $\bs R' \in \mathcal{F}(C_1)$ such that $\|\bs R - \bs R'\| \leq \delta$, where $\mathcal
                F(\cdot)$ denotes the dominant face of a capacity region as in Definition \ref{dom_face}.

                Assume $R$ is a vertex of $C_2$. Then the polymatroid structure of $C_2$ implies
                that $R$ is the intersection of $M$ constraints corresponding to a chain of subsets
                of $\mathcal{M}$. Hence, there is some $k \in \mathcal{M}$ such that $ R_k = f(\{k\}) + \delta
                $. Choose $\bs R'$ as follows

                 \begin{equation}\label{R'_R}
                   R'_i = \left\{ \begin{array}{ll}
                   R_i - \delta, & \textrm{$i = k$}\\
                   R_i, & \textrm{otherwise.}
                   \end{array} \right.
                 \end{equation}

                $\bs R'$ is obviously in a $\delta$-neighborhood of $\bs R$. Moreover, the
                constraint corresponding to the set $\mathcal{M}$ is active for $\bs R'$, so we
                just need to show that $R'$ is feasible in order to prove that it is on the
                dominant face. First, let us consider the sets $S$ that contain $k$. We have
                \begin{equation}\label{k_in_S}
                    \sum_{i \in S} R'_i = \sum_{i \in S} R_i - \delta \leq f(S).
                \end{equation}
                Second, consider the case that $k \notin S$.
                \begin{eqnarray}
                  \sum_{i \in S} R'_i &=& \sum_{i \in S \cup \{k\}} R'_i - R_k + \delta \nonumber \\
                   &\leq& f(S \cup \{k\})  + \delta - R_k \nonumber \\
                   &\leq& f(S) + f(\{k\}) + \delta - R_l \nonumber \\
                   &=& f(S). \nonumber
                \end{eqnarray}

                where the first inequality come from (\ref{k_in_S}), and the second inequality is
                valid because of the submodularity of the function $f(\cdot)$.

               The previous argument establishes that the claim is true for each vertex $\bs R_j$ of the dominant face. But
                every other point $\bs R$ on the dominant face can be represented as
                a convex combination of the vertices, i.e.,
                $$ \bs R = \sum_j \alpha_j \bs R_j, \qquad  \sum_j \alpha_j = 1, \alpha_j \geq 0.$$
                 Using the convexity of the norm function, it is quite straightforward to show that the desired $\bs R'$
                 is given by
                $$ \bs R' = \sum_j \alpha_j \bs R'_j,$$
                where $\bs R'_j$ is obtained for each $\bs R_j$ in the same manner as in
                (\ref{R'_R}).

                 So we have verified that there exists a point, $\bs R$, on the dominant face of $C_1 = C_g(\bs P, \bs H_1)$ such
                 that $\|\bs R_2^* - \bs R\| \leq \delta$. By monotonicity of the utility function the
                 optimal solution $\bs R_2^*$ lies on the dominant face of $C_g(\bs P, \bs H_2)$,
                 hence, from the hypothesis and the fact that $u(\bs R_2^*) \geq u(\bs R_1^*) \geq
                 u(\bs R)$, we conclude
                \begin{equation}\label{opt_dist1}
                    u(\bs R_2^*) - u(\bs R) = |u(\bs R_2^*) - u(\bs R)| \leq B\|\bs R_2^* - \bs R\| \leq B
                    \delta.
                \end{equation}
                Now suppose that $\|\bs R_1^* - \bs R\| > (\frac{B}{A} \delta)^{\frac{1}{2}}$. By the hypothesis in (\ref{AB_hyp}) we
                can write
                \begin{equation}\label{opt_dist2}
                    u(\bs R_1^*) - u(\bs R) = |u(\bs R_1^*) - u(\bs R)| >  B \delta.
                \end{equation}
                By subtracting (\ref{opt_dist1}) from (\ref{opt_dist2}) we obtain $u(\bs R_2^*) <
                u(\bs R_1^*)$ which is a contradiction. Therefore, $\|\bs R_1^* - \bs R\| \leq
                (\frac{B}{A} \delta)^{\frac{1}{2}}$, and the desired result follows immediately by invoking the
                triangle inequality.
        \end{proof}

            The following theorem combines the results of the above lemmas to obtain a bound on the performance
            difference of the greedy and optimal policy.
       \begin{Bound2}\label{Bound2}
            Let $\bs R^*$ be the optimal solution to (\ref{RAC_npctrl}), and $\bar{\mathcal{R}}(\cdot)$ the greedy
            rate allocation policy as defined in (\ref{R_greedy}) for a non-negative concave utility function $u(\cdot)$. Then for every $\epsilon \in
            (0,1]$,
            \begin{eqnarray}\label{Bound2_eps}
                u(\bs{R}^*) &\!\!\!- u\big(\mathbb{E}_{\bs H}\big[\bar{\mathcal{R}}(\bs H)\big]\big)& \leq \epsilon  u(\bs{R}^*) +
                (1-\epsilon)B(\epsilon) \nonumber \\
                &&\Big[\Big(\frac{\sigma_H}{\sqrt{\epsilon}}\Big)^{\frac{1}{2}}+\Big(\frac{B(\epsilon)}{A(\epsilon)}\Big)^{\frac{1}{2}}\Big]\Big(\frac{\sigma_H}{\sqrt{\epsilon}}\Big)^{\frac{1}{2}},
                \nonumber \\
            \end{eqnarray}
            where $B(\epsilon)$ and $A(\epsilon)$ are positive functions of
            $\epsilon$, such that for all $\bs H$ with $d_H(C_g(\bs P,
            \bs H), C_a(\bs P)) ) \leq \frac{\sigma_H}{\sqrt{\epsilon}}$, they satisfy the
            following conditions.
            \begin{eqnarray}\label{Bound2_hyp1}
              |u(\bs{R_a}) - u(\bs{R_g})| &\leq& B(\epsilon) \|\bs{R_a} - \bs{R_g}\|, \nonumber \\ &&
              \foral  \bs{R_a} \in \mathcal{F}(C_a(\bs P)), \nonumber \\
              &&\foral  \bs{R_g} \in \mathcal{F}(C_g(\bs P, \bs H)), \nonumber \\
              \\ \label{Bound2_hyp}
              |u(\bar{\mathcal{R}}(\bs H)) - u(\bs{R})| &\geq& A(\epsilon) \|\bar{\mathcal{R}}(\bs H) - \bs{R}\|^2, \nonumber \\&& \foral \  \bs{R} \in C_g(\bs P, \bs
              H). \\ \nonumber
            \end{eqnarray}

        \end{Bound2}

        \begin{proof}
            Pick any $\epsilon \in (0,1]$. Define the event $\mathcal{V}$ as
            $$d_H(C_g(\bs P, \bs H), C_a(\bs P)) ) \leq \frac{\sigma_H}{\sqrt{\epsilon}}.$$
              By Lemma \ref{region_chebyshev}, the probability of this event is at least
              $1-\epsilon$. Conditioned on $\mathcal{V}$,  we have the following
              \begin{eqnarray}\label{Bound2_ch0}
               \big|u(\bs{R}^*) -u\big(\bar{\mathcal{R}}(\bs H)\big) \big| \leq&&  \!\!\!\!\!\!\!\!\!\!\! B(\epsilon) \|\bar{\mathcal{R}}(\bs H) - \bs{R}^*\| \nonumber \\
               \leq&& \!\!\!\!\!\!\!\!\!\!\!          B(\epsilon)\Big[\Big(\frac{\sigma_H}{\sqrt{\epsilon}}\Big)^{\frac{1}{2}}+\Big(\frac{B(\epsilon)}{A(\epsilon)}\Big)^{\frac{1}{2}}\Big]\Big(\frac{\sigma_H}{\sqrt{\epsilon}}\Big)^{\frac{1}{2}},
               \nonumber \\
              \end{eqnarray}

              where the first inequality follows from  monotonicity of the utility function and (\ref{Bound2_hyp1}). The second  inequality is a direct result of applying Lemma \ref{opt_dist}.

            Using Jensen's inequality as in (\ref{jensen}) we can bound the left-hand side of
            (\ref{Bound2_eps}) as follows
            \begin{eqnarray}\label{bound2_ch1}
              &&u(\bs{R}^*) - u\big(\mathbb{E}_{\bs H}\big[\bar{\mathcal{R}}(\bs H)\big]\big) \nonumber \\
              &&\qquad \leq u(\bs{R}^*) - \mathbb{E}_{\bs H}\big[u\big(\bar{\mathcal{R}}(\bs H)\big)\big] \nonumber \\
              &&\qquad \leq u(\bs{R}^*) - (1-\epsilon)\mathbb{E}_{\bs H}\Big[u(\bar{\mathcal{R}}(\bs H))\Big|
              \mathcal{V}\Big] \nonumber \\
                && \quad \qquad - \mathbf{Pr}(\mathcal{V}^c)\mathbb{E}_{\bs H}\Big[u(\bar{\mathcal{R}}(\bs H))\Big| \mathcal{V}^c\Big]  \nonumber \\
               &&\qquad \leq \epsilon u(\bs{R}^*) + (1-\epsilon)\mathbb{E}_{\bs H}\Big[ |u(\bs{R}^*) -u(\bar{\mathcal{R}}(\bs H)) | \Big| \mathcal{V}
               \Big]. \nonumber \\
            \end{eqnarray}
            In the above relations, the second inequality follows from  $\mathbf{Pr}(\mathcal{V})
            \geq 1- \epsilon$, and the third inequality is obtained from non-negativity of the
            utility function after some manipulation. Replacing (\ref{Bound2_ch0}) in
            (\ref{bound2_ch1}) gives the desired upperbound.
        \end{proof}

\begin{figure}
  \centering
  \includegraphics[width=.5\textwidth]{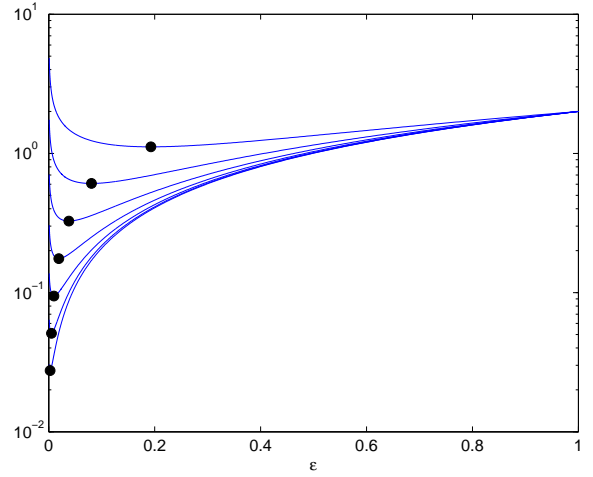}\\
  \caption{Parameterized upperbound on performance difference between greedy and optimal policies.}\label{Bound2_fig}
\end{figure}

        Theorem \ref{Bound2} provides a bound parameterized by $\epsilon$. For very small channel
        variations, $\sigma_H$ tends to zero, and we can choose $\epsilon$ proportional to
        $\sigma_H$ such that the bound in (\ref{Bound2_eps}) approaches zero. Figure
        \ref{Bound2_fig} illustrates the behavior of the parameterized bound provided in
        (\ref{Bound2_eps}) for different values of $\sigma_H$. For each value of $\sigma_H$, the
        upperbound is minimized for a specific choice of $\epsilon$, which is illustrated as a dot
        in Figure \ref{Bound2_fig}. As demonstrated in the figure, for smaller channel variations
        tighter bound is achieved and the minimizer parameter decreases.

        The next theorem
        provides another bound  demonstrating the impact of the structure of the utility
        function on the performance of the greedy policy.

        \begin{Bound1}\label{Bound1}
            Let $\bs R^*$ be the optimal solution to (\ref{RAC_npctrl}) for the non-negative utility function $u(\bs R)$. Also let ${\mathcal{R}^*}(\cdot)$ and ${\bar{\mathcal{R}}}(\cdot)$ be the
            optimal and greedy rate allocation policies, respectively. Then for every $\epsilon \in (0,1]$,
            \begin{equation}\label{Bound1_eps}
                u(\bs{R}^*) - u\big(\mathbb{E}_{\bs H}\big[\bar{\mathcal{R}}(\bs H)\big]\big) \leq \epsilon
                u(\bs{R}^*) +
                \frac{1}{2}(1-\epsilon) r(\epsilon)^2 \Omega,
            \end{equation}
            where $\Omega$ satisfies the following
            \begin{equation}\label{bound2_hyp1}
                \lambda_{\max}(- \nabla^2 u(\bs \xi)) \leq \Omega, \qquad \textrm{for all}\  \bs \xi, \|\bs \xi - \bs
                R^*\| \leq r(\epsilon),
            \end{equation}
            and $r(\epsilon)$ is given by
            \begin{eqnarray}\label{bound1_hyp2}
                &&\!\!\!\!\!\!\!\!\!\!r(\epsilon) =  \sqrt{M}\frac{\sigma_H}{\sqrt{\epsilon}} + \nonumber \\
                && \left[ \sum_{i=1}^M \mathbb{E}_{\bs H}\left[\frac{1}{2} \log \left(\frac{(1+H_i P_i)(1+\sum_{j \neq i} H_j P_j)}{1+\sum_{j=1}^M H_j P_j}\right)\right]^2
                \right]^\frac{1}{2}. \nonumber \\
            \end{eqnarray}
        \end{Bound1}
            \begin{proof}
                Pick any $\epsilon \in (0,1]$. Define the event $\mathcal{V}$ similarly to the proof of
                Theorem \ref{Bound2}.
                Because of monotonicity of the utility function, we know that $\bs R^*$ lies on the dominant face of $C_a(\bs P)$. Since the region $C_a(\bs
                P)$ is the average of all regions $C_g(\bs P, \bs H)$, the optimal policy ${\mathcal{R}^*}(\bs
                H)$ should give a point on the dominant face of $C_g(\bs P, \bs H)$, for almost all $\bs H$. Therefore, conditioned on $\mathcal{V}$, we can
                bound the set in which ${\mathcal{R}^*}(\bs H)$ ranges, i.e., $\|{\mathcal{R}}^*(\bs H) - \bs R^*\| \leq
                r(\epsilon)$, after some straightforward manipulations.
                Now let us write the Taylor expansion of the function $u(\cdot)$ at $\bs R^*$. We
                have
                \begin{eqnarray}
                  u(\bs R) &=& u(\bs R^*) + \nabla u(\bs R^*)'(\bs R - \bs R^*) \nonumber \\
                  &&- \frac{1}{2}(\bs R - \bs R^*)'(-\nabla^2u(\bs \xi))(\bs R - \bs R^*) \nonumber \\
                  &\geq& u(\bs R^*) + \nabla u(\bs R^*)'(\bs R - \bs R^*) \nonumber \\
                  &&- \frac{1}{2}\|\bs R - \bs R^*\|^2 \lambda_{\max}(-\nabla^2u(\bs \xi)), \nonumber \\
                  && \qquad \qquad  \textrm{for some} \  \bs \xi, \|\bs \xi - \bs R^*\| \leq \|\bs R -\bs
                  R^*\|. \nonumber \\
                \end{eqnarray}
                By replacing $\bs R$ by $\mathcal{R}^*(\bs H)$ and conditioning on $\mathcal{V}$ we
                have the following
                \begin{eqnarray}
                   u(\mathcal{R}^*(\bs H))&\geq& u(\bs R^*) + \nonumber \\
                   && \nabla u(\bs R^*)'( \mathcal{R}^*(\bs H)- \bs R^*)
                   - \frac{1}{2} r(\epsilon)^2 \Omega. \nonumber
                \end{eqnarray}
                Now we can bound the left-hand side of (\ref{Bound1_eps}) by bounding the Jensen's
                difference $u(\bs{R}^*) - \mathbb{E}_{\bs H}[u(\mathcal{R}^*(\bs H))]$. After some manipulation similar to
                (\ref{bound2_ch1}), we have
              \begin{eqnarray}\label{bound1_ch1}
              &&u(\bs{R}^*) - u\big(\mathbb{E}_{\bs H}\big[\bar{\mathcal{R}}(\bs H)\big]\big) \nonumber \\
              &&\qquad \qquad\qquad \leq \ u(\bs{R}^*) - \mathbb{E}_{\bs H}\big[u\big(\mathcal{R}^*(\bs H)\big)\big] \nonumber \\
               &&\qquad\qquad\qquad  \leq \ u(\bs{R}^*) - (1-\epsilon)\mathbb{E}_{\bs H}\Big[u({\mathcal{R}^*}(\bs H))\Big| \mathcal{V}\Big] \nonumber \\
               && \qquad\qquad\qquad\quad - \mathbf{Pr}(\mathcal{V}^c)\mathbb{E}_{\bs H}\Big[u({\mathcal{R}^*}(\bs H))\Big| \mathcal{V}^c\Big] \nonumber \\
               &&\qquad\qquad\qquad \leq\  u(\bs{R}^*) - (1-\epsilon)\Big(u(\bs R^*) - \frac{1}{2} r(\epsilon)^2 \Omega \Big) \nonumber \\
               &&\qquad\qquad\qquad = \ \epsilon u(\bs{R}^*) + \frac{1}{2}(1-\epsilon) r(\epsilon)^2 \Omega. \nonumber
            \end{eqnarray}
            \end{proof}

            Similarly to Theorem \ref{Bound2}, Theorem \ref{Bound1} provides a bound parameterized by $\epsilon$ which goes to zero for proper choice of $\epsilon$ as $\Omega$
            becomes smaller and the utility function tends to have a more linear structure. The
            behavior of this parameterized upperbound is also similar to the one illustrated in Figure
            \ref{Bound2_fig}.

            In summary, the performance difference between the greedy and the optimal policy is
            bounded from above by the minimum of the bounds provided by Theorem \ref{Bound2} and
            Theorem \ref{Bound1}.

\section{Conclusion}
We addressed the problem of optimal resource allocation in a fading multiple access channel from an
information theoretic point of view. We formulated the problem as a utility maximization problem
for a more general class of utility functions.

We considered two different scenarios. First, we assumed that the transmitters have power control
feature and the channel statistics are known a priori. In this case, the optimal rate and power
allocation policies are obtained by greedily maximizing a properly defined linear utility function.

In the second scenario, power control and channel statistics are not available. In this case, the
greedy policy is not optimal for nonlinear utility functions. However, we showed that its
performance in terms of the utility is not arbitrarily worse compared to the optimal policy, by
bounding their performance difference. The provided bound tends to zero as the channel variations
become small or the utility function behaves more linearly.

The greedy policy may itself be computationally expensive. A computationally efficient algorithm
can be employed to allocate rates close to the ones allocated by the greedy policy. This algorithm
just takes one iteration of the gradient projection method at each time slot. Under slow fading
conditions, it can be shown that this method tracks the greedy policy very closely, and its
performance is close to the optimal policy.

\bibliographystyle{unsrt}
\bibliography{MAC}

\begin{thebibliography}{10}

\bibitem{TDMA}
X.~Wang and G.B. Giannakis.
\newblock Energy-efficient resource allocation in time division multiple-access
  over fading channels.
\newblock Preprint, 2005.

\bibitem{CDMA1}
S.J. Oh, Z.~Danlu, and K.M. Wasserman.
\newblock Optimal resource allocation in multiservice {CDMA} networks.
\newblock {\em IEEE Transactions on Wireless Communications}, 2(4):811--821,
  2003.

\bibitem{CDMA3}
J.B. Kim and M.L. Honig.
\newblock Resource allocation for multiple classes of {DS-CDMA} traffic.
\newblock {\em IEEE Transactions on Vehicular Technology}, 49(2):506--519,
  2000.

\bibitem{Tse}
D.~Tse and S.~Hanly.
\newblock Multiaccess fading channels part {I}: Polymatroid structure, optimal
  resource allocation and throughput capacities.
\newblock {\em IEEE Transactions on Information Theory}, 44(7):2796--2815,
  1998.

\bibitem{She95}
S.~Shenker.
\newblock Fundamental design issues for the future internet.
\newblock {\em IEEE Journal on Selected Areas in Communications}, 13:1176--118,
  1995.

\bibitem{Srikant}
R.~Srikant.
\newblock {\em Mathematics of {I}nternet Congestion Control}.
\newblock Birkhauser, 2004.

\bibitem{Vishwanath}
S.~Vishwanath, S.A. Jafar, and A.~Goldsmith.
\newblock Optimum power and rate allocation strategies for multiple access
  fading channels.
\newblock In {\em Proceedings of IEEE VTC}, 2001.

\bibitem{power_min}
D.~Yu and J.M. Cioffi.
\newblock Iterative water-filling for optimal resource allocation in {OFDM}
  multiple-access and broadcast channels.
\newblock In {\em Proceedings of IEEE GLOBECOM}, 2006.

\bibitem{QoS}
K.~Seong, R.~Narasimhan, and J.~Cioffi.
\newblock Scheduling for fading multiple access channels with heterogeneous
  {QoS} constraints.
\newblock In {\em Proceedings of International Symposium on Information
  Theory}, 2007.

\bibitem{cover}
T.M. Cover and J.A. Thomas.
\newblock {\em Elements of Information Theory}.
\newblock John Wiley and Sons, Inc., New York, New York, 1991.

\bibitem{Shamai}
S.~Shamai and A.D. Wyner.
\newblock Information theoretic considerations for symmetric, cellular,
  multiple-access fading channels part {I}.
\newblock {\em IEEE Transactions on Information Theory}, 43(6):1877--1894,
  1997.

\bibitem{nlp}
D.P. Bertsekas.
\newblock {\em Nonlinear Programming}.
\newblock Athena Scientific, Cambridge, Massachusetts, 1999.

\bibitem{tech_report}
Ali ParandehGheibi.
\newblock Resource allocation in multiple-access channels.
\newblock {LIDS} technical report, Massachusetts Institute of Technology, 2008.

\end{thebibliography}

\end{document}